# An Efficient Nonlinear Beamformer Based on $P^{th}$ Root of Detected Signals for Linear-Array Photoacoustic Tomography: Application to Sentinel Lymph Node Imaging

**Moein Mozaffarzadeh**[a,b], **Vijitha Periyasamy**[c], **Manojit Pramanik**[c,*], **Bahador Makkiabadi**[a,d,*]

[a]Research Center for Biomedical Technologies and Robotics (RCBTR), Institute for Advanced Medical Technologies (IAMT), Tehran, Iran.
[b]Department of Biomedical Engineering, Tarbiat Modares University, Tehran, Iran.
[c]School of Chemical and Biomedical Engineering, Nanyang Technological University, 62 Nanyang Drive, Singapore.
[d]Department of Medical Physics and Biomedical Engineering, School of Medicine, Tehran University of Medical Sciences, Tehran, Iran.

**Abstract.** In linear-array transducer based photoacoustic (PA) imaging, B-scan PA images are formed using the raw channel PA signals. Delay-and-Sum (DAS) is the most prevalent algorithm due to its simple implementation, but it leads to low quality images. Delay-Multiply-and-Sum (DMAS) provides a higher image quality in comparison with DAS while it imposes a computational burden of $O(M^2)$. In this work, we introduce a nonlinear (NL) beamformer for linear-array PA imaging, which uses the $p^{th}$ root of the detected signals and imposes the complexity of DAS ($O(M)$). The proposed algorithm is evaluated numerically and experimentally (wire-target and *in vivo* sentinel lymph node (SLN) imaging), and the effects of the parameter $p$ are investigated. The results show that the NL algorithm, using a root of $p$ (NL_p), leads to lower sidelobes and higher signal-to-noise ratio (SNR) compared to DAS and DMAS, for ($p > 2$). The sidelobes level (for the wire-target phantom), at the depth of 11.4 $mm$, are about -31 $dB$, -52 $dB$, -52 $dB$, -67 $dB$, -88 $dB$ and -109 $dB$, for DAS, DMAS, NL_2, NL_3, NL_4 and NL_5, respectively, indicating the superiority of the NL_p algorithm. In addition, the best value of $p$ for SLN imaging is reported to be 12.

**Keywords:** Photoacoustic imaging, nonlinear beamformer, linear-array imaging, noise suppression, contrast improvement..

*Manojit Pramanik, manojit@ntu.edu.sg
*Bahador Makkiabadi, b-makkiabadi@tums.ac.ir

## 1 Introduction

Photoacoustic/Optoacoustic imaging (PAI) is an emerging medical imaging technique combining optical contrast and spatial resolution of ultrasound (US).[1,2] PAI is based on the photoacoustic (PA) effect which provide structural, functional and potentially the molecular information of tissue.[3,4] In PAI, a laser pulse irradiates the sample/tissue, resulting in local temperature rise, and due to thermoelastic expansion, pressure waves (in the form of ultrasound waves) are generated.[5] The ultrasound waves (also known as photoacoustic wave) travels within the medium and then are recorded using US transducer. These PA waves are used to obtain the optical absorption map of the inside of the tissue.[6] In circular geometry photoacoustic tomography (PAT), a single-element/ring array US transducer can be used to acquire the PA signals around the tissue in full 360°.[3,7] Recently, low-cost PAT systems are extensively being investigated where pulsed laser diodes are used to make the PA systems more compact, portable, and affordable.[8,9] However, such imaging system based on single-element/ring array ultrasound transducer is difficult to translate into clinical applications.[10] Therefore, recently dual-modal clinical ultrasound and photoacoustic imaging have been reported for better clinical translation.[11,12]

Clinical ultrasound systems work with linear, convex or phased array transducers. Here, in this work we focus on one of the clinical applications of sentinel lymph node (SLN) imaging, where linear array ultrasound probe is usually used.[13] Sentinel lymph node biopsy (SLNB) is a standard clinical procedure done in breast cancer staging. SLNB replaced axillary lymph node dissection (ALND) where the lymph nodes around the tumour is removed. Removal of nodes is necessary to curb metastasis of cancerous cells. ALND is an unwanted procedure in node-negative patients and it leads to side-effects such as lymphedema, arm weakness and infections of breast. Hence, in SLNB, biopsy samples of the lymph nodes to which the mammary glands drain first are examined before removing the nodes.[14,15] SLNs are identified by intradermal injection of dyes (such as methylene blue).[16] Since, the injected methylene blue dye gives a strong PA signal, feasibility of PA imaging for SLN identification has been experimented widely on small animals and in patient trials.[13,17–19]

One of the crucial challenges in linear-array based PAT is the image reconstruction.[20–22] The presence of noise and artifacts in the detected signals and simplifications of the reconstruction algorithms for speed degrade the PA image quality.[23] Indeed, there are some inherent artifacts, caused by the image formation algorithms, observed in the reconstructed images.[24] Moreover, due to the limited view of the linear-array transducer (in comparison with the ring-array transducer, which get a full 360 degree view) reconstructed image quality is poor.[10,11] In other words, linear-array transducers have a view of approximately only 40°, and the image quality is lower compared to circular tomography.[25–27]

Beamforming algorithms, which are commonly used in Radar and US imaging, can be used in PAT with some modifications.[28] There are some studies conducted to use a single beamforming algorithm for the integrated US/PA imaging systems.[25,29,30] Delay-and-Sum (DAS) can be considered as the most common beamforming algorithm in US and PAT.[31–33] To address the in-capabilities in DAS, which is mainly due to its blindness, Minimum Variance (MV) can be used.[34,35] In MV, all the calculated samples for each point of imaging are weighted adaptively, resulting in a significant resolution, but it should be noticed that the sidelobes will not be well-degraded. On the other hand, Delay-Multiply-and-Sum (DMAS) can be used to suppress the noise and sidelobes, and improve the image quality.[36] To address the low noise suppression of DMAS at the presence of high level of noise, Double-Stage DMAS (DS-DMAS), in which two stages of DMAS is used, is recently introduced for linear-array US and PAT.[37,38] Despite all the improvement gained by DMAS and DS-DMAS, the resolution improvement is not satisfying, compared to MV. The matter of low contrast of MV and low resolution of DMAS is addressed using the MV-Based DMAS (MVB-DMAS) where the combination of these two methods are used for PA image reconstruction.[28,39] Eigenspace-Based MV (EIBMV) is introduced for US image formation to degrade the sidelobes and improve the contrast obtained by the MV.[40,41] The concept of MVB-DMAS is applied to the EIBMV and used for linear-array PAT.[42] Coherence factor (CF) is applied to the MV beamformed signals to improve the resolution and suppress the sidelobes.[43] Two modifications of CF are intro-duced for linear-array PAT in order to have a lower sidelobes and higher resolution, compared to the conventional CF.[44,45]

In this work, a novel image formation algorithm for linear-array PAT is proposed. The proposed algorithm is a Nonlinear (NL) beamformer, and it uses the $p^{th}$ root of the detected PA signals. The main improvement gained by the proposed method is lower sidelobes and noise while the complexity of the algorithm is in the order of DAS ($O(M)$). The results show (both numerical and experimental) that the proposed algorithm can be an appropriate choice for linear-array PA image formation, especially when a high level of noise affect the image quality.

## 2 Methods and Materials

### 2.1 Beamforming

Assuming a linear geometry for PA waves detection, the optical absorption distribution of the tissue can be reconstructed using DAS which can be written as follows:

$$y_{DAS}(k) = \sum_{i=1}^{M} x_i(k - \Delta_i), \quad (1)$$

where $y_{DAS}(k)$, $k$, $M$ are the output of the beamformer, the time index and the number of elements of the array, respectively. In addition, $x_i(k)$ and $\Delta_i$ are the detected signals and the corresponding time delay for the detector $i$, respectively. As mentioned before, DAS results in a low quality image and is commonly used due to its simple implementation. To improve the contrast gained by DAS,

DMAS is introduced,[36] which is as follows:

$$y_{DMAS}(k) = \sum_{i=1}^{M-1} \sum_{j=i+1}^{M} x_i(k-\Delta_i)x_j(k-\Delta_j). \quad (2)$$

The dimensionally squared problem of (2) is addressed as follows:[36]

$$x'_{ij}(k) = \text{sign}[x_i(k-\Delta_i)x_j(k-\Delta_j)]\sqrt{|x_i(k-\Delta_i)x_j(k-\Delta_j)|}. \quad (3)$$

$$y_{DMAS}(k) = \sum_{i=1}^{M-1} \sum_{j=i+1}^{M} x'_{ij}(k). \quad (4)$$

The new components appear in the spectrum due to the similar ranges of frequency for $x_i(k-\Delta_i)$ and $x_j(k-\Delta_j)$, and the multiplication procedure in the DMAS algorithm. A band-pass filter is applied on the beamformed output signal to only pass the necessary frequency components, while keeping the one centered on $2f_0$ almost unaltered. As a results of the used filter, it is named Filtered-DMAS (F-DMAS), extensively evaluated in the reference.[36] In the previous publications, the superiority of the DMAS has been proved in the terms of sidelobes, resolution and contrast. However, all the advantages are achieved at the expense of a higher computational burden. DMAS is a nonlinear beamformer as a results of its correlation procedure. In this paper, it is proposed to use the $p^{th}$ root of the detected signals to improve the contrast of the PA images.[46] The $p^{th}$ root of the signals are used as the input of the DAS algorithms. The proposed method formula can be written as follows:

$$y_{NL\_p}(k) = \left(\frac{1}{M}\sum_{i=1}^{M} \text{sign}(x_i(k-\Delta_i))\sqrt[p]{|x_i(k-\Delta_i)|}\right)^p \quad (5)$$

As can be seen in the (5), the $p^{th}$ root of the detected signals are used in the summation procedure. The same as DMAS algorithms, which changes the dimension to the order of $Volt^2$, the dimension of the signals would be changes to $Volt^{1/p}$ in the NL algorithm. This problem is addressed by the ($p^{th}$) power finally used (bringing back the dimension to the $Volt$). It should be noticed that for $p = 2$, the NL algorithm would be a close approximation of the DMAS beamformer. To illustrate this, consider the expansion of (5) when $p = 2$:

$$y_{NL\_2}(k) = \left(\frac{1}{M}\sum_{i=1}^{M} \text{sign}(x_i(k-\Delta_i)) \sqrt[2]{|x_i(k-\Delta_i)|}\right)^2 = \left(\frac{1}{M}\sum_{i=1}^{M} x'_i(k)\right)^2 =$$
$$\left(\frac{1}{M}\sum_{i=1}^{M} x'_i(k)\right)\left(\frac{1}{M}\sum_{i=1}^{M} x'_i(k)\right) = \qquad (6)$$
$$\frac{1}{M^2}\left(x'_1(k) + x'_2(k) + ... + x'_M(k)\right)\left(x'_1(k) + x'_2(k) + ... + x'_M(k)\right).$$

By multiplying the term in the parentheses in the (6), the following equation is generated:

$$\frac{1}{M^2}\Bigg[\left(\boxed{x'^2_1(k)} + x'_1(k)x'_2(k) + x'_1(k)x'_3(k) + ... + x'_1(k)x'_M(k)\right) +$$
$$\left(x'_2(k)x'_1(k) + \boxed{x'^2_2(k)} + x'_1(k)x'_3(k) + ... + x'_1(k)x'_M(k)\right) + \qquad (7)$$
$$\left(x'_M(k)x'_1(k) + x'_M(k)x'_2(k) + ... + x'_M(k)x'_{M-1}(k) + \boxed{x'^2_M(k)}\right)\Bigg].$$

As shown in the circles in (7), there are some terms with the power of 2 (the terms inside the circles). All the term in the circles can be written as follows:

$$\frac{1}{M^2}\left(x'^2_1(k) + x'^2_2(k) + ... + x'^2_M(k)\right) = \frac{1}{M^2}\underbrace{\sum_{i=1}^{M} x'^2_i(k)}_{\text{a DAS on the }p^{th}\text{ power of the signals}} \qquad (8)$$

where, as is clarified in the equation, there is a DAS implemented on the $p^{th}$ power of the signals. The rest of the terms of (7), those who are not in the (8), can be mathematically written as follows:

$$\frac{2}{M^2} \underbrace{\sum_{i=1}^{M-1} \sum_{j=i+1}^{M} x'_i(k) x'_j(k)}_{\text{DMAS}} \quad (9)$$

Finally, the (6) is mathematically written as follows:

$$\frac{1}{M^2} \underbrace{\sum_{i=1}^{M} x'^2_i(k)}_{\text{a DAS on the } p^{th} \text{ power of the signals}} + \frac{2}{M^2} \underbrace{\sum_{i=1}^{M-1} \sum_{j=i+1}^{M} x'_i(k) x'_j(k)}_{\text{DMAS}} \quad (10)$$

As can be seen, (6) has led to two terms (a DAS and a DMAS). The results in the next section show that (6) would be a close approximation of DMAS. In fact, the performance of DMAS has been achieved while only the computational complexity of DAS has been expended. A flow-chart is presented in Fig. 1 to better illustrate the proposed reconstruction method. It should be noticed that for even p-values, the sign of the PA signals will be lost by the final power of p, leading to the splitting of the original spectrum into DC components and doubled spectrum band of the original signal. Therefore, at the end of procedure of image formation, a band-pass filter should be applied on the beamformed PA signals to only pass the necessary frequency components. In the next section, the numerical and experimental results of the proposed method for linear-array PAT are reported.

## 2.2 Numerical Study

The K-wave Matlab toolbox is used to carry out the numerical study.[47] Six pairs of point targets, having a radius of 0.1 $mm$, are positioned at the depth of 25 $mm$ until 50 $mm$, separated 5 $mm$ in axial axis and 4 $mm$ in lateral axis, and two single-point targets are positioned at the depth of 32.5 $mm$ and 42.5 $mm$. A linear-array having $M$=128 elements operating at 4 $MHz$ central frequency and 77% fractional bandwidth is used to detect the PA signals. The speed of sound is assumed to be 1540 $m/s$ during simulations. The sampling frequency is 50 $MHz$.

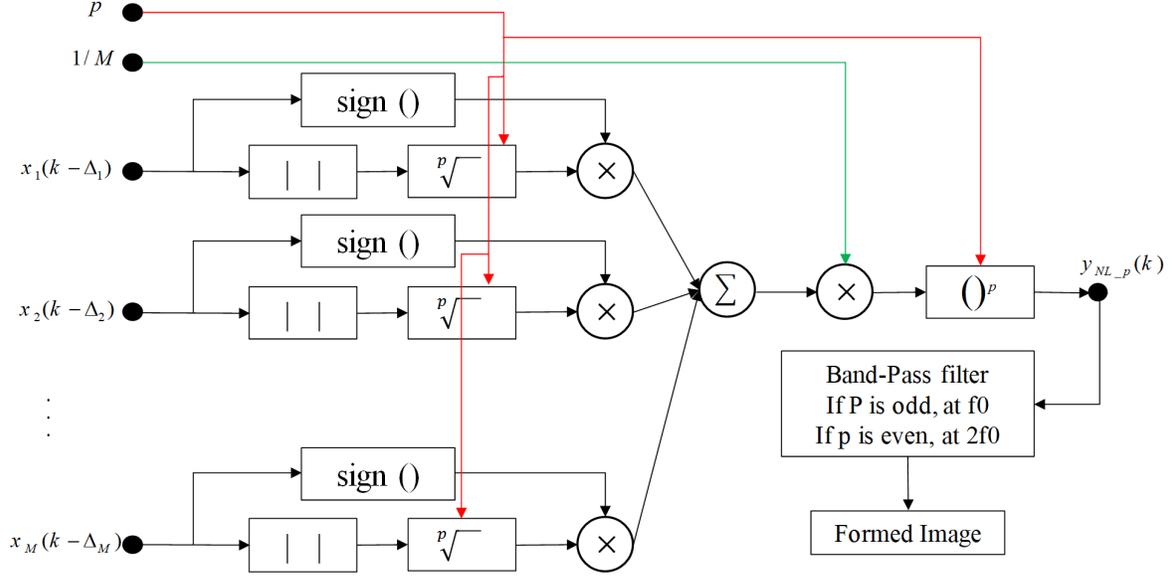

Fig 1: The flow-chart of the proposed NL beamformer.

*2.3 Wire-Target Phantom Experiment*

Experimental PA images were acquired using the clinical ultrasound system (E-CUBE 12R, Alpinion, South Korea). The ultrasound transducer has 128 elements over a length of 3.85 $cm \times 1$ $cm$, center frequency of 8.5 $MHz$ and the fractional bandwidth is 95%. Laser beam of 532 $nm$ from Nd:YAG pump laser (Continuum, Surelite Ex, San Jose, California, USA) is used for excitation. The laser has a pulse repetition rate of 10 $Hz$.[13,48,49] Lens of focal length -50 $mm$ is used to diverge the laser beam to illuminate the target which is four pencil leads of diameter 0.5 $mm$ positioned over an area of 25 $mm \times$ 25 $mm$. 1% of the laser beam is reflected to a photodiode which is used as a trigger signal to synchronize the ultrasound system and the laser excitation.[50] For every trig-ger the ultrasound system acquires channel data from 64 elements of the array transducer. Hence, the PA images are acquired at a frame rate of 5 $frames/s$. Acquired radio frequency data were saved in the local machine and used later for testing the reconstruction algorithm. The pencil leads (Mountain peak, China) were pinned in clay and immersed in water. Laser passed through the wall of water tank before reaching the target. Leads were at a depth of 5 $mm$ to 15 $mm$ from the transducer.

*2.4 In vivo Imaging of Sentinel Lymph Node*

The experimental set-up for the *in vivo* imaging of sentinel lymph node is shown in Fig. 2. Laser beam of 1064 $nm$ (the same as the wire-target phantom experiment) is focused into a fiber bundle using a 150 $mm$ converging lens. A beam-splitter is used to reflect a small percentage of the beam to photodiode which is needed to trigger the clinical ultrasound system. The fiber bundle has 1600 multimode fibers which are separated into two bundles at the output. The two bundles are held across the ultrasound transducer using a custom-designed 3-D printed holder. The angle of illumination from both the fiber bundles are 15°. Adult Sprague Dawley rat of 250 $gms$ is initially anesthetized using a cocktile of ketamine (85 $mg/kg$) and xylazine (15 $mg/kg$). The hair in the region of interest is depleted and then inhalation anesthesia of 1 $L/min$ oxygen and 0.75% isoflurane (Euthanex Corp.), is given. The fur on the scalp of the animal is removed using a hair clipper. The hair removal cream (Veet, Reckitt Benckiser) is gently applied to the shaved area for further depletion of the fur. The applied cream is removed after 4-5 $min$ using a cotton swab. A chicken breast tissue of 5 $mm$ thickness is placed on the animal to mimic SLN imaging of human. Black ink is injected in the forepaw of the animal and massaged for the ink to flow into the lymph node (Fig. 2(b)). The region of interest is illuminated with 20 $mJ/cm^2$, of energy which is within the maximum permissible limit of 100 $mJ/cm^2$ for 1064 $nm$ as per American National Standard for Safe Use of Lasers.[51] After the acquisition of PA images the animal is euthanized with overdose of pentobarbital. An incision is made to expose the SLN (Fig. 2(c)). The SLN is excised (Fig. 2(d)) from the animal. All experiments are performed in accordance with the guidelines and regulations approved by the Institutional Animal Care and Use Committee of Nanyang Technological University, Singapore (Animal Protocol Number ARF-SBS/NIE-A0263).

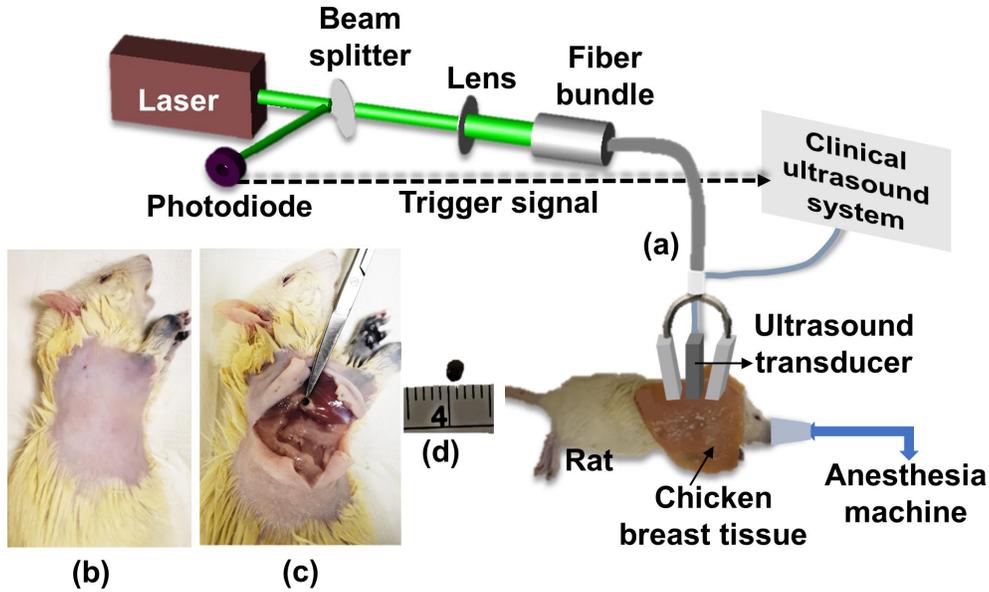

Fig 2: Experimental set-up of *in vivo* imaging, (a) experimental set-up, (b) photograph of rat used for *in vivo* imaging, (c) photograph of the rat exposing the sentinel lymph node after incision, (d) excised sentinel lymph node stained black.

## 3 Results

### 3.1 Numerical Study

First, the performance of the proposed method is evaluated for $p = 2$ and $p = 3$, compared to DAS and DMAS. The reconstructed PA images are shown in Fig. 3 where noise was added to the detected signals considering a SNR of 30 $dB$. As can be seen in Fig. 3(a), DAS results in high sidelobes, which degrades the image quality, and the effects of the added noise are obvious in the image. DMAS reduces the sidelobes and artifacts, but they still affect the image. NL_2 provides an image the same as DMAS while NL_3 provides a higher noise and sidelobes suppression, which leads to a higher image quality in comparison with other mentioned beamformers. To evaluate the performance of the beamformers in details, the lateral variations of the images shown in Fig. 3 are presented in Fig. 4. The same performance of NL_2 and DMAS are clear considering their lateral variations, which are almost the same, while they outperform DAS in the terms of sidelobes and lateral valley. Considering Fig. 4(a), it can be seen that level of sidelobes for DAS, DMAS and NL_2 are about -25 $dB$, -37 $dB$ and -37 $dB$, respectively. On the other hand, NL_3 provides a lower sidelobes and lateral valley. As a result, NL_3 degrades the sidelobes for about 21 $dB$, 9 $dB$ and 9 $dB$, compared to DAS, DMAS and NL_2, respectively. We have calculated full-width-half-maximum (FWHM) to evaluate the lateral resolution of the beamformers. At the depth of 40 $mm$, DAS, DMAS, NL_2 and NL_3 lead to a FWHM of 2.05 $mm$, 1.43 $mm$, 1.43 $mm$ and 1.17 $mm$, respectively. Thus, NL_3 results in lower FWHM, indicating a better resolution, while it is not a significant improvement compared to those obtained in previous studies.[28,39] These results indicate that NL method can be used to provide a high image quality.

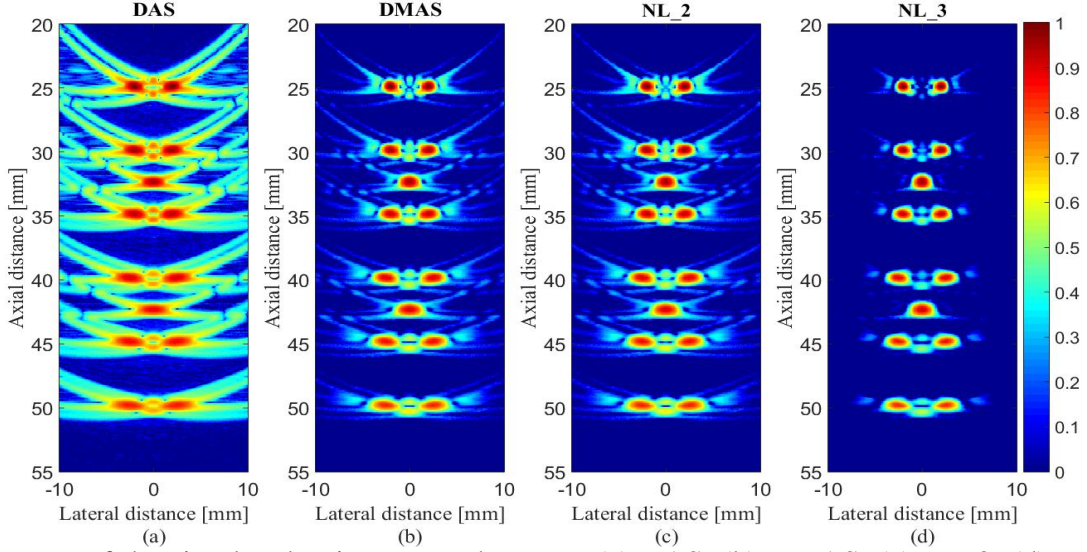

Fig 3: Images of the simulated point-target phantom. (a) DAS, (b) DMAS, (c) NL_2, (d) NL_3. All the images are shown with a dynamic range of 60 $dB$. Noise was added to the detected signals considering a SNR of 30 $dB$.

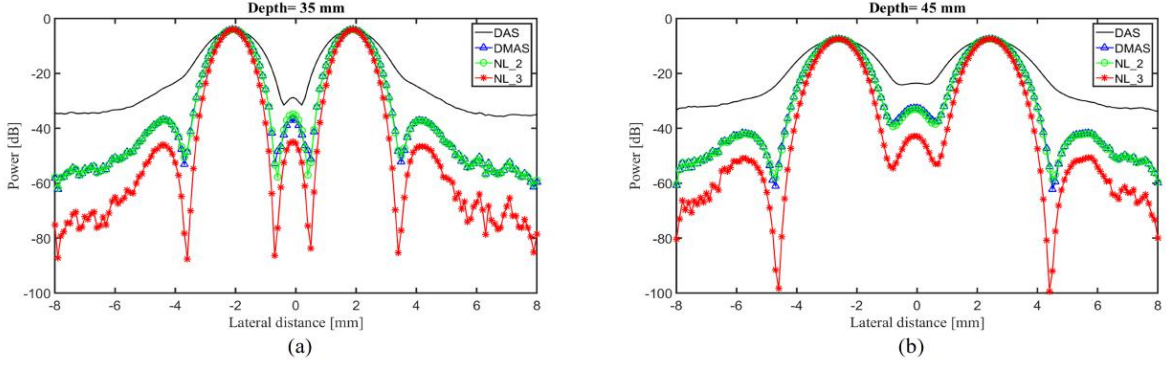

Fig 4: Lateral variations of DAS, DMAS, NL_2 and NL_3 at the depths of (a) 35 $mm$ and (b) 45 $mm$ using the images shown in Fig. 3.

*3.1.1 High Level of Medium Noise*

To further evaluate the NL beamformer, noise is added to the detected signals considering a SNR of 0 $dB$, and the reconstructed image are shown in Fig. 5. The presence of high power noise clearly reduces the quality of the image obtained by DAS while the effects are mitigated using DMAS and NL_2. Considering the image shown in Figs. 5(b) and 5(c), it is demonstrated that the noise still degrades the image quality. However, the effects of the noise are highly suppressed using NL_3, as shown in Fig. 5(d). The lower sidelobes, noise and lateral valley, gained by the NL_3, at the presence of the powerful noise, are illustrated Fig. 6 using the lateral variations. To quantitatively compare the methods, the Signal to Noise Ratio (SNR) has been used, which is as follows:

$$SNR = 20\log_{10} P_{signal}/P_{noise}, \tag{11}$$

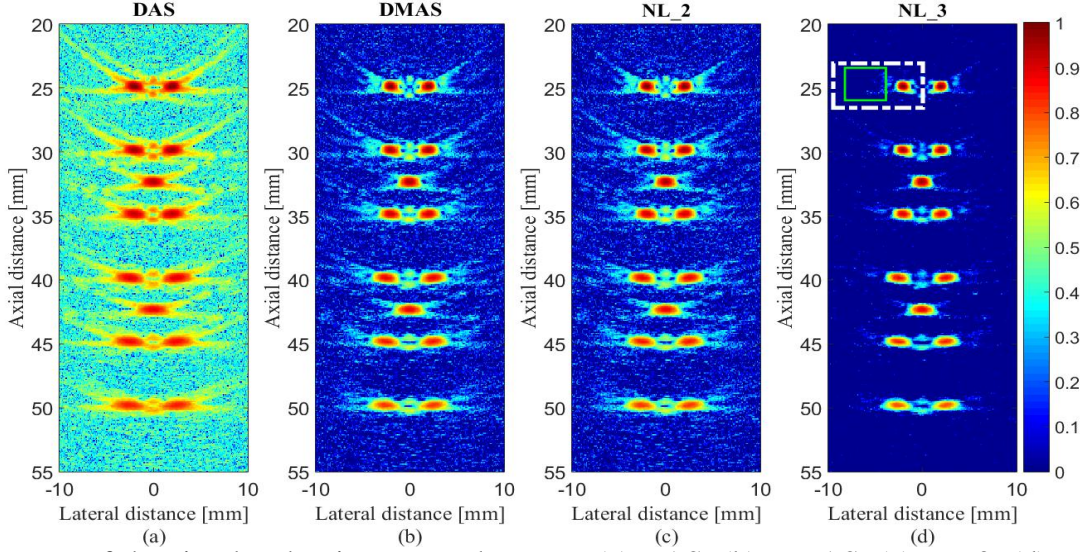

Fig 5: Images of the simulated point-target phantom. (a) DAS, (b) DMAS, (c) NL_2, (d) NL_3. All the images are shown with a dynamic range of 60 $dB$. Noise was added to the detected signals considering a SNR of 0 $dB$.

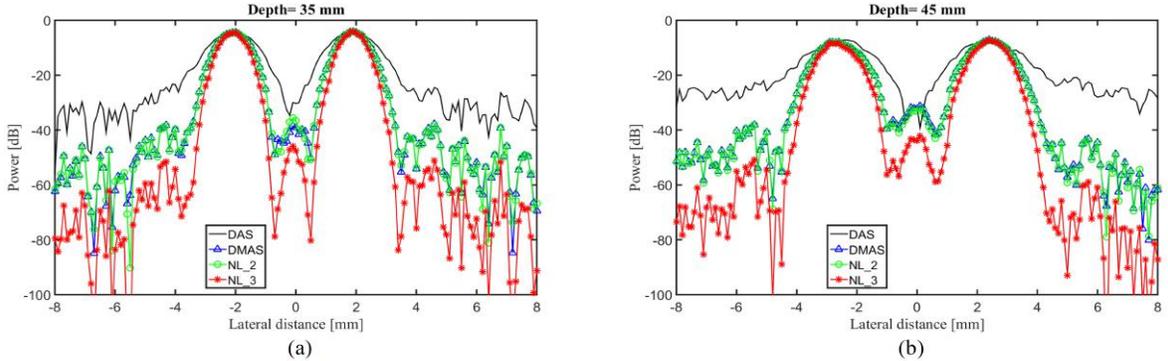

Fig 6: Lateral variations of DAS, DMAS, NL_2 and NL_3 at the depths of (a) 35 $mm$ and (b) 45 $mm$ using the images shown in Fig. 5.

where $P_{signal}$ and $P_{noise}$ are difference of maximum and minimum intensity of a rectangular region including a point target (white dashed rectangle in Fig. 5(d)), and standard deviation of the noisy part of the region (green rectangle in Fig. 5(d)), respectively.[37,38] The calculated SNRs are shown in Table 1, at the all depths of imaging. As demonstrated, the proposed method when $p = 2$ performs almost the same as DMAS while both of them provide a higher SNR compared to DAS. This is mainly due to higher noise and sidelobes suppression. It should be noticed that the same performance is achieved while the computational burden of the NL_2 algorithm is of the order of magnitude of M ($O(M)$), which is lower than DMAS with ($O(M^2)$). On the other hand, increasing the root used in the proposed method to 3 (NL_3) would result in a higher noise suppression which leads to a higher SNR. Consider for instance, the depth of 45 $mm$, where DAS, DMAS, NL_2 and NL_3 result in a SNR of about 23.71 $dB$, 32.07 $dB$, 32.17 $dB$ and 38.98 $dB$, respectively. In other word, NL_3 improves the SNR for about 15.27 $dB$, 6.91 $dB$ and 6.91 $dB$ in comparison with DAS, DMAS and NL_2, respectively.

Table 1: SNR ($dB$) values at the different depths.

| Depth($mm$) | DAS   | DMAS  | NL_2  | NL_3  |
|-------------|-------|-------|-------|-------|
| 25          | 32.85 | 46.50 | 46.52 | 59.33 |
| 30          | 30.50 | 44.74 | 44.80 | 57.01 |
| 35          | 29.70 | 43.30 | 43.36 | 54.96 |
| 40          | 25.17 | 35.5  | 35.69 | 44.65 |
| 45          | 23.71 | 32.07 | 32.17 | 38.98 |
| 50          | 21.39 | 28.55 | 28.64 | 34.81 |

*3.1.2 Effects of the parameter P*

Here, we evaluate the effects of $p$ on the reconstructed PA images. The reconstructed images using different $P$s are presented in Fig. 7 where increasing the amount of $p$ would improve the image quality and degrade the sidelobes. The further evaluation can be conducted using the lateral variations shown in Fig. 8 where the bigger the amount of $p$, the lower the sidelobes. It can be seen that in each step of increasing $p$, the sidelobes are degraded for about 13 $dB$. The FWHM improvement is not significant when $p$ is increased while the SNR improvement would be considerable since the sidelobes are degraded. Considering the presented numerical results, one might come to conclusion that increasing the parameter $p$ would always result in image improvement. However, in the section 4, the effect of parameter $p$ is investigated using experimental data, and it is shown that a large or wrong $p$ would result in image disturbance and signal removal. In other words, when it comes to select the best $p$ (providing the highest SNR), we need to consider the main signal removal of the NL beamformer as well as the noise/artifacts removal.

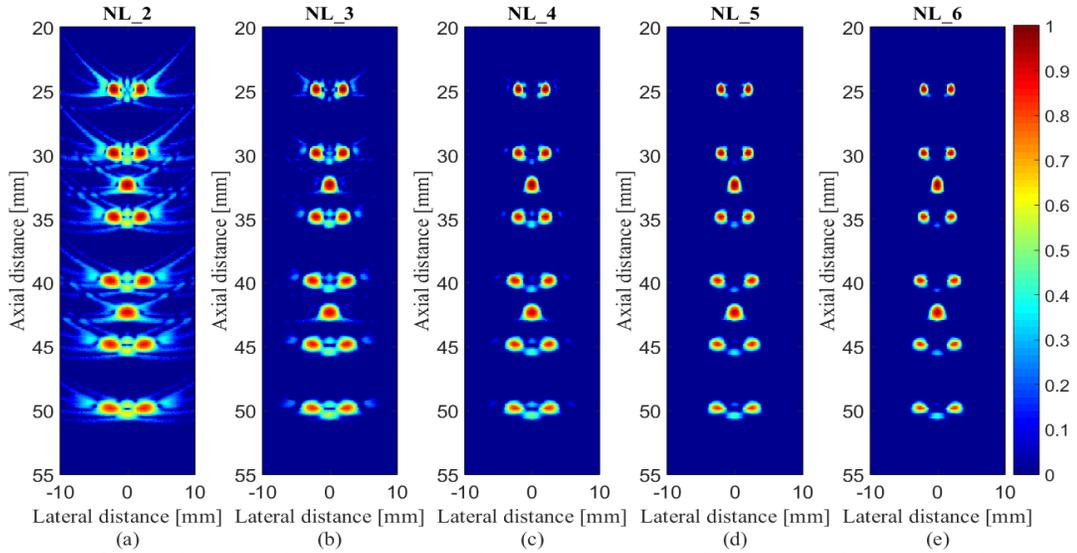

Fig 7: Images of the simulated point-target phantom. (a) NL_2, (b) NL_3, (c) NL_4, (d) NL_5 and (e) NL_6. All the images are shown with a dynamic range of 60 $dB$. Noise was added to the detected signals considering a SNR of 30 $dB$.

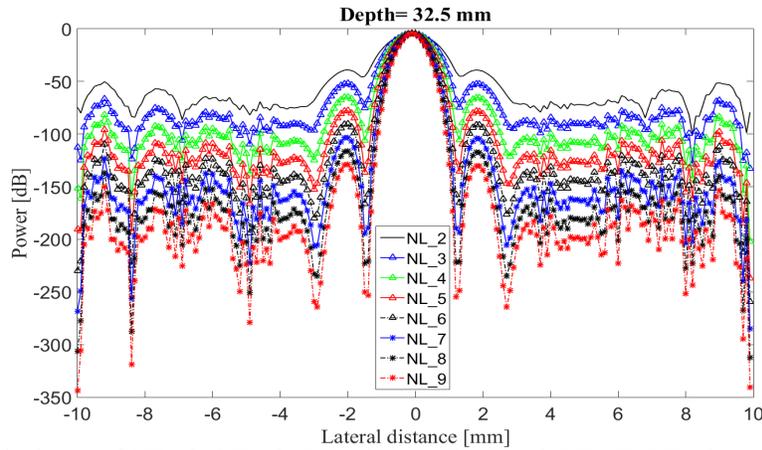

Fig 8: Lateral variations of NL_2, NL_3, NL_4, NL_5, NL_6, NL_7, NL_8 and NL_9 at the depths of 32.5 $mm$ using the images shown in Fig. 7.

### 3.2 Wire-Target Phantom Experiment

The reconstructed experimental images are shown in Fig. 9 (zoomed version in Fig. 10) where DAS leads to a low quality image having a high sidelobes, artifacts and noise. DMAS and NL_2 results in the same image quality (Figs. 9(b) and 9(c)), but increasing the root of the NL beamformer leads to image quality improvement, as demonstrated in (Figs. 9(e) and 9(f)). To have a better look at the gained improvement, consider the lateral variations presented in Fig. 11 where the NL_5 method outperforms DAS, DMAS and NL with lower $p$ (see the circle and arrows). It should be noticed that, as mentioned in the section 2, the performance of the DMAS and NL_2 are almost the same, which is also proved with the experimental data, considering their lateral variations. To quantitatively evaluate the NL beamformer, we have calculated the SNR for the experimental PA images. The SNRs are presented in Table 2 where, at the depth of 11.3 $mm$ NL_5 improves the SNR for about 19.1 $dB$, 14.5 $dB$, 14.5 $dB$, 9.8 $dB$ and 4.5 $dB$, compared to DAS, DMAS, NL_2, NL_3 and NL_4, respectively. All the experimental results indicate the superiority of the NL beamformer compared to DAS and DMAS.

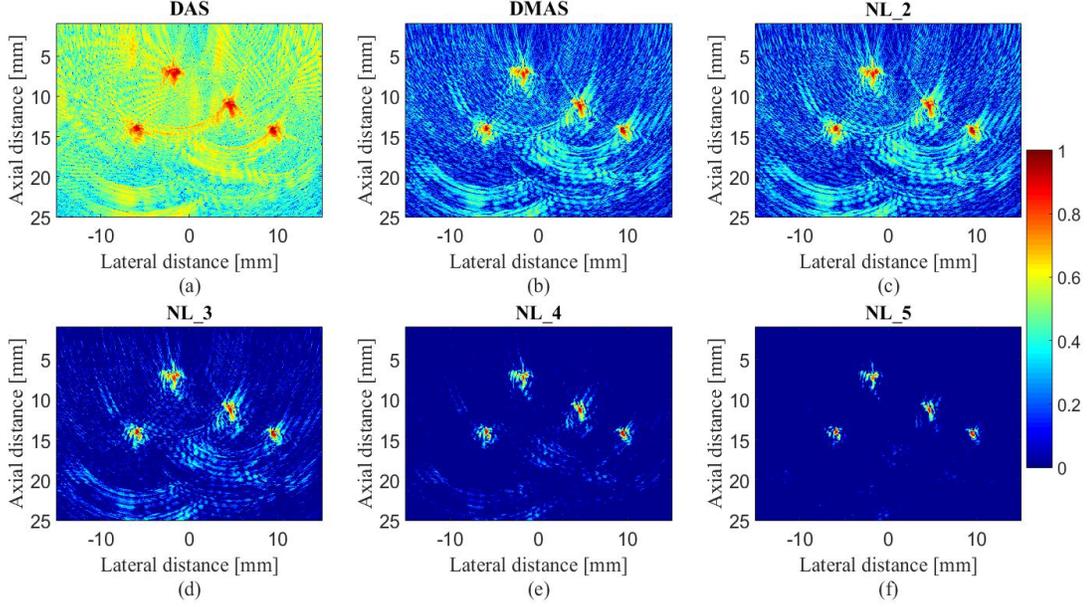

Fig 9: The reconstructed experimental images using (a) DAS, (b) DMAS, (b) NL_2, (c) NL_3, (d) NL_4 and (e) NL_5. A wire-target phantom was utilized. All the images are shown with a dynamic range of 70 $dB$.

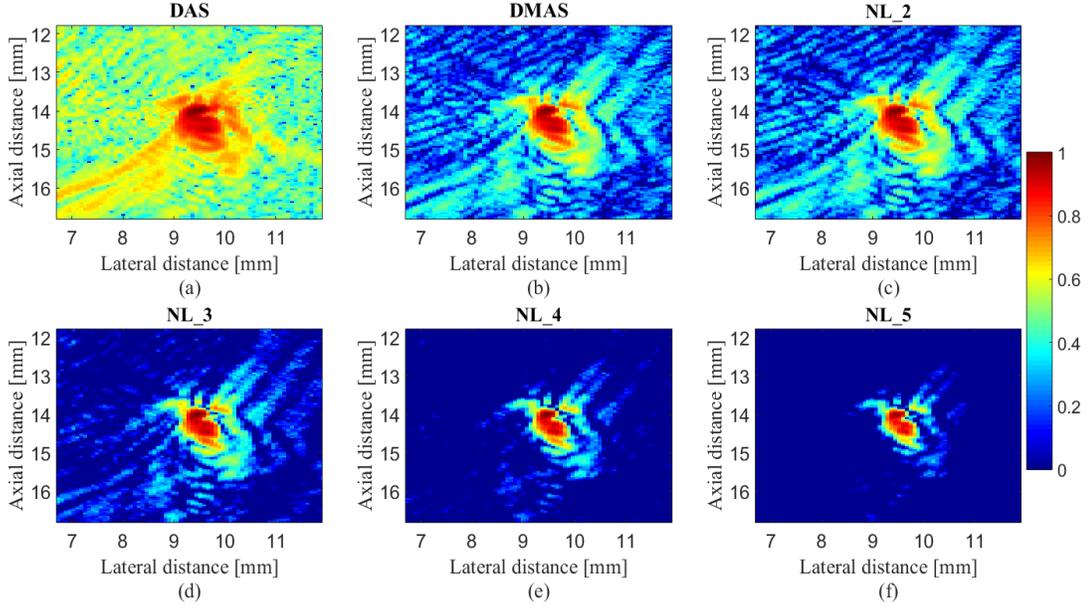

Fig 10: A zoomed version of the Fig. 9 for better comparison.

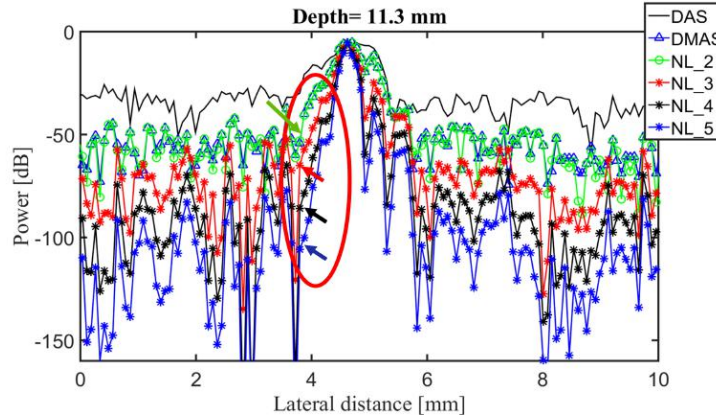

Fig 11: The lateral variations for the PA images shown in Fig. 9.

## 3.3 In vivo Imaging of Sentinel Lymph Node

The reconstructed *in vivo* images are demonstrated in Fig. 12 where the place the SLN has been indicated with the arrow (see Fig. 12(c)). As can be seen in Fig. 12(f), the background noise of the image reconstructed by the NL_5 is better suppressed in comparison with other images. In other words, the NL beamformer (for $p > 2$) provides higher image quality compared to DAS and DMAS. It should be mentioned that at the wavelength of 1064 $nm$, the blood vessels have a very poor contrast. Therefore, they do not show up or get buried in the background.

Table 2: SNR ($dB$) values for the experimental image shown in Fig. 9.

| Depth($mm$) | DAS | DMAS | NL_2 | NL_3 | NL_4 | NL_5 |
|---|---|---|---|---|---|---|
| 7.1 | 35.8 | 44.0 | 44.0 | 50.7 | 56.4 | 59.9 |
| 11.3 | 24.8 | 29.4 | 29.4 | 34.1 | 39.4 | 43.9 |

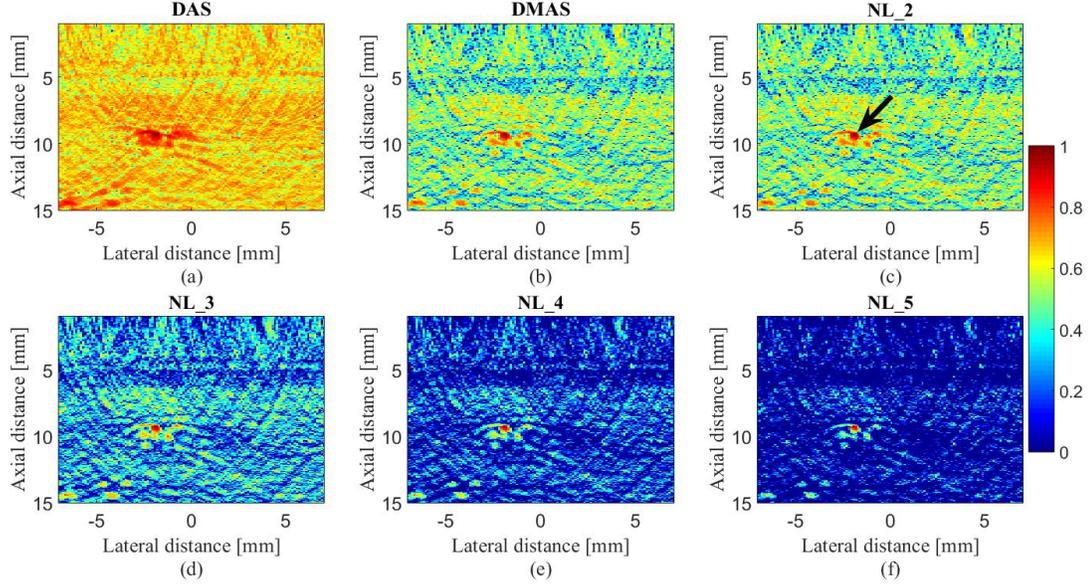

Fig 12: The reconstructed *in vivo* images using (a) DAS, (b) DMAS, (b) NL_2, (c) NL_3, (d) NL_4 and (e) NL_5. All the images are shown with a dynamic range of 60 $dB$.

## 4 Discussion

In this work, we have introduced a novel image reconstruction algorithm (NL beamformer) for PAT in the case that a linear-array US transducer has been used for data acquisition. We have extensively evaluated the proposed method for Photoacoustic imaging in the terms of SNR, FWHM, level of sidelobes and target separability. In addition, we have evaluated the effects of the parameter p on the reconstructed images, and employed the proposed method for SLN imaging, for the first time. The DAS algorithm considers all the signals the same which is why it results in a low quality image having a high sidelobes and artifacts. The DMAS beamformer uses a correlation procedure to suppress the noise and the sidelobes caused by the simple summation in DAS beamformer, and finally results in higher image quality compared to DAS. The main improvements gained by the proposed method are the lower sidelobes and a higher SNR, compared to DMAS, while a lower computational burden is imposed. As can be seen in the Fig. 3 and Fig. 5, the NL beamformer results in a high image quality, especially when $p$ is increased. This is because of the suppression of the weak signals while the powerful signals are maintained. To put it more simply, using the $p^{th}$ root of the detected PA signals suppresses the artifacts and noise. While the powerful samples are maintained, the contribution of the weak samples of the detected PA signals would be lower. It is expected to detect the targets using the powerful samples, and the $p^{th}$ root makes it possible. The lateral variations shown in Fig. 4 and Fig. 6, where the proposed method has been evaluated at the presence of high levels of imaging noise and medium heterogeneity, prove the superiority of the NL algorithm. Table 1 shows the higher noise suppression of the NL algorithm which, as was mentioned, results from the usage of the $p^{th}$ root. Despite all the promising numerical results, it was necessary to evaluate the NL beamformer with experimental data. Fig. 9 and Fig. 11 show the reconstructed PA image when a wire-target phantom is used. The experimental results indicate that the NL beamformer (with $p > 2$) outperforms DAS and DMAS, which makes it an appropriate algorithm for PA image reconstruction. The superiority of the NL beamformer is also evaluated with an *in vivo* experiment, and the results are shown in Fig. 12. To pass the necessary information and attenuate the DC component generated after the NL_p (with an even p), a band-pass filter was applied by a Tukey window ($\alpha$=0.5) to the beamformed PA signal spectra, covering 4.5-11.5 $MHz$ and 11-19 $MHz$ for numerical and experimental studies, respectively. The algorithm is independent of the laser wavelength is used. Thus, in experiments, there is no reason to use only a single wavelength for evaluation. At 1064 $nm$, the black ink has a strong optical absorption, so, it is used as the contrast agent. Any other contrast agents having strong absorption can be used.

The matter of selection of $p$ is of importance in the NL beamformer. As mentioned, using the $p^{th}$

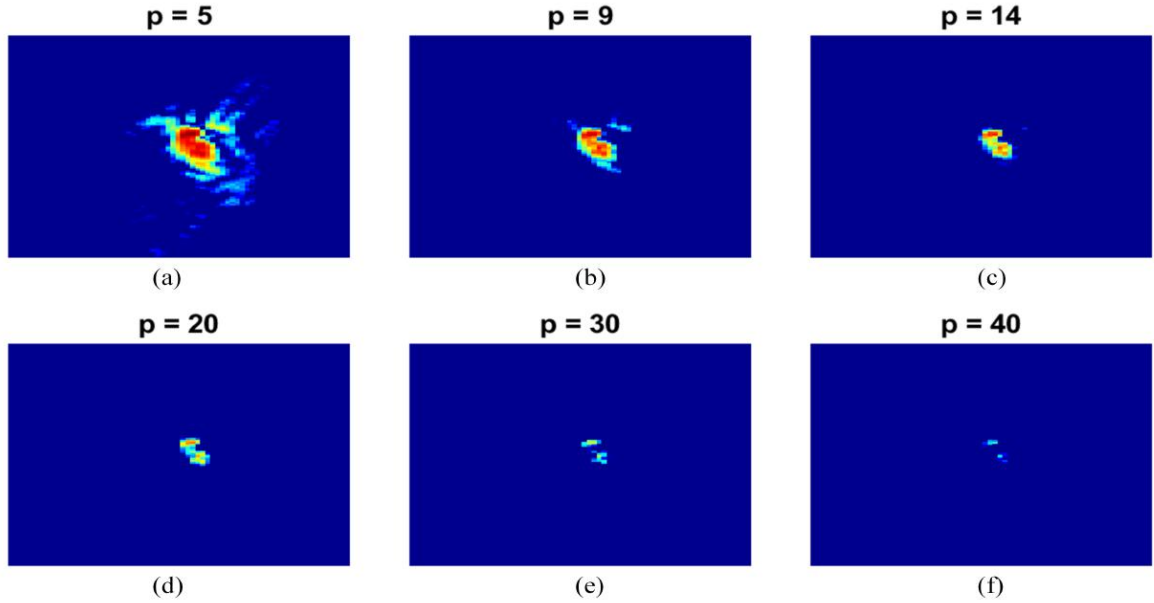

Fig 13: The effects of $p$ on the experimental images shown in Fig. 10.

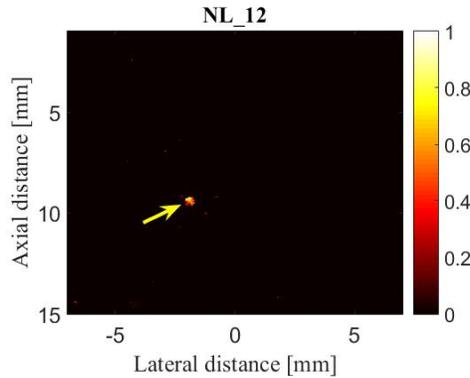

Fig 14: The reconstructed PA image using NL_12 where the yellow arrow shows the target.

root degrades the artifacts and noise contribution in the procedure of PA image reconstruction, and the numerical results indicated that the larger the parameter $p$, the higher image quality. However, if the parameter $p$ get selected so large (larger than the best $p$), it could suppress the signals caused by the target of imaging as well as the noise and artifacts. To illustrate this, consider Fig. 13 where large $p$ is used for reconstruction as well as a low amount of $p$. It can be seen that having a $p$ of 40, or even 30, would suppress the main target of imaging as well as the background noise and sidelobes. In other words, wrong selection of $p$ may result in a disturbed and nonsense image quality, as can be seen in Figs. 13(e) and 13(f). Here, we show the application of the proposed algorithm for SLN imaging, but the algorithm can be used in other applications as well. Due to our investigation, for SLN imaging, selection of $p = 12$ would result in the best image quality (can be seen in Fig. 14) without compromising the main target of imaging (SLN). It should be noted that $p = 12$ was obtained empirically. An adaptive way of choosing the best $p$, based on the energy or power of the detected PA signals, would be a matter of investigation for our future studies. Beamformers usually can be utilized in the US and PA image formation. The proposed method changes the dynamic range of the detected signals, and we can use it to improve the image quality. The proposed nonlinear reconstruction will certainly affect the quantitative values of the signal, making it ineligible for a quantitative data analysis. However, PAI is not only about multi-spectral imaging. There are other applications such as structural imaging as well. Thus, in the case of structural imaging, the proposed algorithm would be useful. As shown in Fig. 11 and Fig. 8, Even though the pattern of noise is not changed, its level is reduced. We can use this reduction to improve the image quality in PAI. It worth to mention that the proposed method in this paper, especially when a $p$ larger than 3 is selected, may not be an appropriate algorithm for US imaging. This is mainly because of the importance of speckles in US images. To put it more simply, in most of US images, the speckles contain important information for diagnosis and detection. To this end, using $p > 3$ for US imaging is not suggested since it would result in speckle removal which is not desired. In,[36] it has been mentioned that the computational burden of DMAS is in the order of $M^2$ ($O(M^2)$) while the order of DAS is $M$ ($O(M)$). The formula of the NL beamformer is shown in (5), and in section 2, it is proved that NL_2 could be a close approximation of the DMAS algorithm while it only imposes an order of $M$. This is mainly due to the fact that there is just a summation procedure in the proposed method, and all the improvements by NL are achieved while a lower computational burden is imposed, compared to DMAS. Of note, considering the sign and calculation of $p^{th}$ root in the computational complexity, the NL_p algorithm imposes a complexity in the order of $M * log(M)$. The size of the experimental images is $256 \times 256$ pixel where DAS, DMAS and NL_p take a time of 0.16 $sec.$, 17.9 $sec.$ and 0.26 $sec.$ to reconstruct the images, respectively. In addition, the size of numerical images is $550 \times 200$ pixel where DAS, DMAS and NL_p take a time of 0.18 $sec.$, 31.69 $sec.$ and 0.40 $sec.$ to reconstruct the images, respectively. It deserves to mention that the proposed method can be implemented on a processing chip such as FPGA and be used in the investigation and commercial PA devices. As mentioned above, the selection of parameter $p$ highly affects the performance of the proposed method, especially when it becomes to choose the best $p$. Thus, implementation of the proposed algorithm on a FPGA and designing an adaptive way to select the parameter $p$ can be considered as the further works.

## 5 Conclusion

In this work, a novel image formation algorithm (NL_p) was utilized for linear-array PAT. While DMAS improves the image quality, compared to DAS, it imposes a computational burden of $O(M^2)$. The introduced method uses the $p^{th}$ root of the detected signals, and a summation, resulting in $O(M)$. The NL algorithm was evaluated numerically and experimentally (wire-target and *in vivo* SLN imaging). The results indicated that NL outperforms DMAS while it imposes the computational burden of DAS. The reconstructed experimental images of the wire-target phantom showed that, at the depth of 11.3 $mm$, NL_5 leads to SNR improvement of about 77%, 49%, 49%, 28% and 11% in comparison with DAS, DMAS, NL_2, NL_3 and NL_4, respectively. In addition, the appropriate $p$ for SLN imaging was obtained to be 12.

**Funding**

The research is supported by Tier 2 grant funded by the Ministry of Education in Singapore (ARC2/15: M4020238 to M.P.).

**Acknowledgments**

The authors, VP and MP acknowledge Dr. Dienzo Rhonnie Austria for his help in animal experiments.

**Disclosures**

Authors have no relevant financial interests in the manuscript and no other potential conflicts of interest to disclose.

ulation and experimental validation," *Journal of biomedical optics* **22**(4), 041008–041008 (2017).

51 *American National Standards Institute (ANSI). Laser Institute of America, Orlando, FL* , Z136.1 (2007).

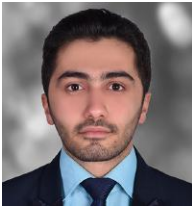 **Moein Mozaffarzadeh** was born in Sari, Iran, in 1993. He received the B.Sc. degree in Electrical Engineering from Babol Noshirvani University of Technology (Mazandaran, Iran), in 2015, and the M.Sc. degree in Biomedical Engineering from Tarbiat Modares University (Tehran, Iran), in 2017. He is currently a research assistant at research center for biomedical technologies and robotics, institute for advanced medical technologies (Tehran, Iran). His current research interests include Photoacoustic Image Reconstruction, Ultrasound Beamforming and Biomedical Imaging.

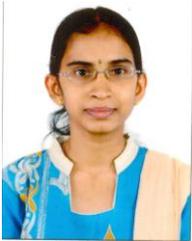 **Vijitha Periyasamy** joined Wayne State University since 2015-May as a PhD student in functional and molecular ultrasound research laboratory (http://ultrasound.eng.wayne.edu/). His interesting areas are Medical Image, Object Detection and Data Mining. He had strong background in Computer Science pattern recognition and computer graphics. He holds two bachelors, Computer science and Post and Telecommunication. He also received a master in computer science from Wayne State university 2017. His current research interests include Endocavity Ultrasound and Photoacoustic for fetal and maternal care.

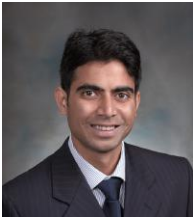 **Manojit Pramanik** received his PhD in biomedical engineering from Washington University in St. Louis, Missouri, USA. He is currently an assistant professor of the School of Chemical and Biomedical Engineering, Nanyang Technological University, Singapore. His research interests include the development of photoacoustic/thermoacoustic imaging systems, image reconstruction methods, clinical application areas such as breast cancer imaging, molecular imaging, contrast agent development, and Monte Carlo simulation of light propagation in biological tissue.

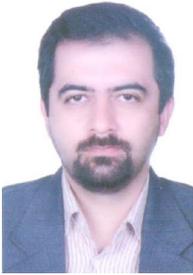 **Bahador Makkiabadi** received his B.Sc. degree in Electronics Engineering from Shiraz University (Shiraz, Iran), the M.Sc. in Biomedical Engineering from Amirkabir University of Technology (Tehran, Iran), and the Ph.D. in Biomedical Engineering from University of Surrey (Guildford, Surrey, UK). Currently, he is working at Research Center for Biomedical Technologies and Robotics (RCBTR), Institute for Advanced Medical Technologies (IAMT), Tehran University of Medical Sciences, Tehran, Iran. His research interests include Blind Source Separation, Advanced Array Signal Processing for Medical Applications and Biomedical Imaging.

# List of Figures

2  Experimental set-up of *in vivo* imaging, (a) experimental set-up, (b) photograph of rat used for *in vivo* imaging, (c) photograph of the rat exposing the sentinel lymph node after incision, (d) excised sentinel lymph node stained black.

3  Images of the simulated point-target phantom. (a) DAS, (b) DMAS, (c) NL_2, (d) NL_3. All the images are shown with a dynamic range of 60 $dB$. Noise was added to the detected signals considering a SNR of 30 $dB$.

4  Lateral variations of DAS, DMAS, NL_2 and NL_3 at the depths of (a) 35 $mm$ and (b) 45 $mm$ using the images shown in Fig. 3.

5  Images of the simulated point-target phantom. (a) DAS, (b) DMAS, (c) NL_2, (d) NL_3. All the images are shown with a dynamic range of 60 $dB$. Noise was added to the detected signals considering a SNR of 0 $dB$.

6  Lateral variations of DAS, DMAS, NL_2 and NL_3 at the depths of (a) 35 $mm$ and (b) 45 $mm$ using the images shown in Fig. 5.

7  Images of the simulated point-target phantom. (a) NL_2, (b) NL_3, (c) NL_4, (d) NL_5 and (e) NL_6. All the images are shown with a dynamic range of 60 $dB$. Noise was added to the detected signals considering a SNR of 30 $dB$.

8  Lateral variations of NL_2, NL_3, NL_4, NL_5, NL_6, NL_7, NL_8 and NL_9 at the depths of 32.5 $mm$ using the images shown in Fig. 7.

9  The reconstructed experimental images using (a) DAS, (b) DMAS, (b) NL_2, (c) NL_3, (d) NL_4 and (e) NL_5. A wire-target phantom was utilized. All the images are shown with a dynamic range of 70 $dB$.

10  A zoomed version of the Fig. 9 for better comparison.

11  The lateral variations for the PA images shown in Fig. 9.

## List of Tables